\documentstyle[twocolumn,eqsecnum,aps]{revtex}
\begin{document}
\draft
%\preprint{HEP/123-qed}
\title{Parity Invariance and Effective Light-Front Hamiltonians}
\author{M. Burkardt}
\address{
Department of Physics\\
New Mexico State University\\
Las Cruces, NM 88003-0001\\
U.S.A.}
%\date{\today}
\maketitle
\begin{abstract}
In the light-front form of field theory, boost invariance
is a manifest symmetry. On the downside, parity and rotational 
invariance are not manifest, leaving the possibility 
that approximations or incorrect renormalization might lead to
violations of these symmetries for physical observables.
In this paper, it is discussed how one can turn this deficiency 
into an advantage and utilize parity violations (or the absence 
thereof) in practice for constraining effective light-front 
Hamiltonians. More precisely, we will identify observables that 
are both sensitive to parity violations and easily calculable 
numerically in a non-perturbative framework and we will use these 
observables to constrain the finite part of non-covariant 
counter-terms in effective light-front Hamiltonians.
\end{abstract}
%\pacs{Valid PACS appear here.
%{\tt$\backslash$\string pacs\{\}} should always be input,
%even if empty.}

\narrowtext

\section{Introduction}
\label{sec:intro}
Light-Front (LF) quantization is very similar to canonical equal
time (ET) quantization \cite{di:49} (here we closely 
follow Ref. \cite{kent}). Both are Hamiltonian formulations of
field theory, where one specifies the fields on a
particular initial surface. The evolution of the fields
off the initial surface is determined by the
Lagrangian equations of motion. The main difference
is the choice of the initial surface, $x^0=0$ for
ET and $x^+=0$ for the LF respectively.
In both frameworks states are expanded in terms of fields
(and their derivatives) on this surface. Therefore,
the same physical state may have very different
wave functions\footnote{By ``wave function'' we mean here
the collection of all Fock space amplitudes.}
in the ET and LF approaches because fields at $x^0=0$
provide a different basis for expanding a state than
fields at $x^+=0$. The reason is that the microscopic
degrees of freedom --- field amplitudes at $x^0=0$
versus field amplitudes at $x^+=0$ --- are in general
quite different from each other in the two formalisms.

From the purely theoretical point of view, various advantages
of LF quantization derive from properties of the ten generators
of the Poincar\'e group (translations $P^\mu$,
rotations ${\vec L}$ and boosts ${\vec K}$) \cite{di:49,kent}.
Those generators which leave the initial surface
invariant (${\vec P}$ and ${\vec L}$ for ET and
$P_-$, ${\vec P}_\perp$, $L_3$ and ${\vec K}$ for LF)
are ``simple'' in the sense that they have very simple
representations in terms of the fields (typically just
sums of single particle operators). The other generators, 
which include the ``Hamiltonians'' ($P_0$, which is conjugate
to $x^0$ in ET and $P_+$, which is conjugate to the LF-time
$x^+$ in LF quantization) contain interactions among the
fields and are typically very complicated.
Generators which leave the initial surface invariant are also
called {\it kinematic} generators, while the others are called
{\it dynamic} generators. Obviously it is advantageous to have as
many of the ten generators kinematic as possible. There are
seven kinematic generators on the LF but only six in ET quantization.

The fact that $P_-$, the generator of $x^-$ translations, is
kinematic (obviously it leaves $x^+=0$ invariant!)
and positive has striking
consequences for the LF vacuum\cite{kent}. For free fields $p^2=m^2$ implies
for the LF energy $p_+ = \left(m^2 + {\vec p}_\perp \right)/2p_-$.
Hence positive energy excitations have positive $p_-$. After the
usual re-interpretation of the negative energy states this implies
that $p_-$ for a single particle is non-negative [which makes sense,
considering that $p_- =\left(p_0-p_3\right)/\sqrt{2}$].
$P_-$ being kinematic
means that it is given by the sum of single particle momenta $p_-$.
Combined with the non-negativity of $p_-$ this implies that, 
even in the presence of interactions,
the physical vacuum (ground state of the theory) differs
from the Fock vacuum (no particle excitations) only by
so-called {\it zero-mode} excitations, i.e. by excitations of modes
which are independent of the longitudinal LF-space coordinate
$x^-$. Due to this simplified vacuum structure, the LF-framework
seems to be the only framework, where a constituent quark picture
in a strongly interacting relativistic field theory has a chance
to make sense \cite{all:lftd,wi:mb,wi:dr,brodsky}.

Whenever the generator of a symmetry is dynamical
(contains interactions) it is somewhere between
very difficult and impossible to monitor and maintain 
that symmetry at each step of a calculation --- unless
of course on can solve the theory exactly.
A typical example is the boost invariance, which, in the context
of equal time quantization, is generated by a dynamical operator.
It is thus {\it not manifestly} true that $E_n^2=m_n^2+{\vec p}^2$,
i.e. the eigenvalues satisfy the correct dispersion relation if and 
only if one starts from the correct renormalized Hamiltonian, with
the right counter-terms (for the regulators employed).
Now suppose, one does not know the Hamiltonian precisely but has
some idea how it may look like: for example one knows the operators
that appear in the Hamiltonian but not their coefficients.
In such a situation it should, at least in principle,\footnote{In practice,
this example has a serious flaw: Most non-perturbative numerical techniques
project most efficiently on the ground state with zero
momentum. Energies of excited states and states with nonzero 
momentum are typically much more difficult to evaluate.}
be possible
to use relativistic covariance as a renormalization condition
that can be used to pin down some of the renormalization
constants in the Hamiltonian\footnote{This is of course
only possible as long as the energy scale of the approximations 
involved is much larger than the kinetic energy associated with 
the momentum $p$.}.

In Hamiltonian LF calculations one faces a very similar problem:
in practical non-perturbative calculations one often leaves 
out degrees of freedom, such as zero-modes and other 
high-energy degrees of freedom \cite{all:lftd}. Because of such
(in practical non-perturbative calculations nearly unavoidable!) 
approximations it is in general not
guaranteed that one recovers non-manifest
symmetries (on the LF: parity and rotational invariance)
in the end. On the contrary, without 
appropriate\footnote{Appropriate means here not only the
correct infinite part of the counter-term, which can often
be obtained on the basis of perturbative arguments, but
also the correct finite part of the counter-term.}
counter-terms in the Hamiltonian one is almost guaranteed to
violate these symmetries.

In this work, an attempt will be made to turn these problems
into an advantage. More precisely, we will identify
physical observables which are easily accessible in a practical
non-perturbative calculation and which are sensitive
to violations of covariance.

Parity transformations take $x^+\!\!\longleftrightarrow \!x^-$, i.e.
LF-time and LF-space are interchanged. Within the LF formalism,
this is a very complicated transformation: in the above language, 
parity transformations are obtained by {\it dynamical}
operators because the initial surface ($x^+=0$) is not
invariant under parity.
From the practical point-of-view, this has the following consequences.
First, given a LF Hamiltonian, most approximations
to that Hamiltonian are likely to break parity invariance.
Secondly, even if one does a ``perfect numerical job'',
parity invariance may still be broken because it may have
been broken already at the level of regularization and
renormalization: most regulators that are practically
useful within the LF-formalism break parity invariance.
This also includes effects that arise when zero-modes
are omitted (or eliminated) as dynamical degrees of freedom
\cite{hari1,hari2,le:ap,ho:vac,dgr,mbsg,brett,fr:eps,mb:adv}.

The counter-terms introduced in the renormalization procedure
are thus not only supposed to cancel the infinities but
also to restore parity invariance (in the limit as the
cutoff goes away). In general, restoring parity requires an 
additional finite renormalization! This issue will
be the main subject in the rest of this paper.

Many of the general statements made so far also apply to
rotational invariance, i.e. at least in principle this paper could
also have been written about rotational invariance \cite{mbrot}!
However, in practical LF calculations, rotational invariance is usually broken
much more badly than parity invariance: for example, in the
transverse lattice formulation of LF field theory
\cite{bardeen,paul,pauldoubl,mb:elfe,mb:conf}, the classical action is still
invariant under parity (which maps the 1+1 dimensional sheets
onto themselves), but not under general rotations which mix the
continuous longitudinal direction and the discrete transverse
direction. Thus for a given transverse lattice (with fixed
transverse spacing), if one does a perfect numerical job and if
one does the renormalization right, one should obtain a
perfectly parity invariant theory, whereas, under the same
conditions, rotational invariance should only be recovered
if one furthermore takes the limit of zero lattice spacing
and infinite lattice volume.

The paper is organized as follows. In Section \ref{sec:parity}
and \ref{sec:sense}, some
observables that are sensitive to violations of parity are 
identified and we will discuss their usefulness in the context
of practical non-perturbative LF-calculations. In Section \ref{sec:nonex}, 
we will illustrate the formalism by studying the these observables 
in the context of a concrete example: 1+1 dimensional
Yukawa theory. In Section \ref{sec:summary}, we will summarize the findings
and discuss potential applications of the formalism to QCD.

\section{The difficulty in finding sensitive and sensible
observables}
\label{sec:parity}
There are of course infinitely many relations between matrix
elements one can write down that are potentially
sensitive to parity violations.
However, most of them are not useful here because of
a number of practical considerations:
The main limitations arise since
\begin{itemize}
\item[A)] certain relations arising from parity invariance
are ``protected'' by some manifest symmetry, or
\item[B)] the matrix elements appearing in those relations are
incalculable in praxis.
\end{itemize}
These points can be illustrated by considering a few examples.
\subsection{Protected Relations}

Charge conjugation is a manifest symmetry in the LF formalism.
Therefore, certain matrix elements that are in principle
sensitive to parity could be ``protected'' by C-parity.
For example, if $|n\rangle $ is an eigenstate of parity then its
vacuum to meson
scalar and pseudoscalar couplings
($\langle 0| \overline{\psi}\psi |n\rangle $ and 
$\langle 0| \overline{\psi}i\gamma_5\psi |n\rangle $ respectively)
cannot both be nonzero for the same state $|n\rangle $. However, the same statement
is true for eigenstates of C-parity --- irrespective whether 
$|n\rangle $ is an eigenstate of parity. Thus, no matter how badly
parity is violated, as long as manifest C-parity is maintained,
either $\langle 0| \overline{\psi}\psi |n\rangle $ or
$\langle 0| \overline{\psi}i\gamma_5\psi |n\rangle $ (or both)
will always vanish and one cannot exploit these vacuum to
meson matrix elements to investigate whether or not parity
is violated.

The situation changes when one introduces different quark flavors
and considers states that have net flavor and hence are not eigenstates
of C-parity, such as $\overline{s}u$\footnote{Of course, for a
flavor symmetric theory G-parity takes the role of C-parity in this case,
but we will assume $m_u \neq m_s$ in this example.}. 
However, even in a multi-flavor theory, these vacuum to meson matrix
elements are not very useful in practice for investigations of
parity invariance because of some nontrivial operator renormalization issues,
which we will discuss below in the context of $j^-$.

\subsection{Incalculable Matrix Elements}

Other selection rules and relations can be derived from the
Lorentz transformation properties of vector currents in a
parity invariant theory. For example, for the vacuum to meson
matrix element of a vector/pseudovector one obtains

\noindent 3+1 dimensions:
\begin{eqnarray}
\langle 0|\overline{\psi}\gamma_\mu \psi |n,s,p\rangle
&=& s_\mu c_n \nonumber\\
\langle 0|\overline{\psi}\gamma_\mu \gamma_5\psi |n,s,p\rangle
&=& p_\mu c_n
\end{eqnarray}
1+1 dimensions:\footnote{Even if one is only interested in 3+1
dimensional theories, it is useful to consider the 1+1 dimensional
relations because the 3+1 dimensional relations assume
rotational invariance and are thus likely to be broken.
Furthermore the transverse lattice formulation of 3+1 dimensional
field theories explicitly utilizes 1+1 dimensional degrees of
freedom to approximate the 3+1 dimensional continuum.}
\begin{eqnarray}
\langle 0|\overline{\psi}\gamma_\mu \psi |n,p\rangle
&=& \varepsilon_{\mu \nu} p^\nu c_n \nonumber\\
\langle 0|\overline{\psi}\gamma_\mu \gamma_5\psi |n,p\rangle
&=& p_\mu c_n,
\end{eqnarray}
where $p_\mu$, $s_\mu$ are the momentum and spin vector 
respectively. Note that there is no spin in 1+1 dimensions.
$\varepsilon_{\mu \nu}$ is the antisymmetric tensor
in 1+1 dimensions. 

Naively one may think that these relations
are useful to detect violations of parity invariance.
For example,one can calculate both
$c_n^{(+)} \equiv \langle 0|\overline{\psi}\gamma^+ \gamma_5\psi
|n,p\rangle /p^+$ and
$c_n^{(-)} \equiv \langle 0|\overline{\psi}\gamma^- \gamma_5\psi
|n,p\rangle /p^-$
independently and then compare the results: in a
parity invariant theory the results for $c_n$ extracted from
the two Lorentz components of the current should be
the same. There are many such relations that one can derive
for matrix elements and coupling constants calculated from
different Lorentz components. For the purpose of detecting
violations of parity invariance in the Hamiltonian
LF formalism they are all totally useless!

The flaw in all these examples is that at least one
of the Lorentz components of the currents 
involved in such relations contains a {\it bad current}:
in the LF formalism one usually decomposes the fermion field
into dynamical and non-dynamical components
\begin{equation}
\psi = \psi_{(+)} + \psi_{(-)},
\end{equation}
where
$\psi_{(\pm)} \equiv \frac{1}{2} \gamma^\mp \gamma^\pm \psi$.
The point is that the (LF-) time derivative of
$\psi_{(-)}$ does not enter the Lagrangian and thus
$\psi_{(-)}$ satisfies a constraint equation and is
usually eliminated by solving this constraint 
equation.\footnote{This procedure resembles very much the
elimination of the `Coulomb component' of the photon field
in Coulomb gauge.} An explicit example will be given
in section \ref{sec:nonex}. Since the solution to these
constraint equations are typically {\it nonlinear}
expressions (in term of the dynamical degrees of freedom),
any operator containing $\psi_{(-)}$ naturally ends up
being highly nonlinear when expressed in terms of
$\psi_{(+)}$ and the other (dynamical) fields involved in
the interactions. For example, in a gauge theory
($A^+=0$ gauge) or Yukawa theory, $\psi_{(-)}$
contains a product of $\psi_{(+)}$ and the boson field.
These nonlinear terms generally lead to nasty divergences
in composite operators involving $\psi_{(-)}$, which is
the reason why composite operators involving $\psi_{(-)}$
are called {\it bad currents}.\footnote{Sometimes one
refers to operator which are bilinear in $\psi_{(-)}$,
such as $j^- = \psi_{(-)}^\dagger\psi_{(-)}$,
as {\it very bad} operators.}
However, the motivation for this terminology is not only the
occurrence of divergences --- after all we have become used 
to divergences in quantum field theory and we have learned
how to renormalize the infinities by adding counter-terms --- 
but the fact that the finite part of the counter-term
remains a priory ambiguous in this procedure.
The bottom line is the following: the goal of this paper
is to find ways to use space-time symmetries to constrain
the finite parts of the non-covariant counter-terms in the
LF-Hamiltonian. Matrix elements of bad currents contain
unknown finite renormalizations themselves. In a sense,
by considering matrix elements of bad currents we have increased
not only the number of equations (renormalization conditions)
but also the number of ``unknowns'' and the net result
is questionable. \footnote{Certain bad currents also enter
the Hamiltonian and thus studying their matrix elements does not
increase the number of unknowns because these operators have
to be renormalized anyway in order to construct the
Hamiltonian. However, we will not exploit this fact here any
further. See Ref. \cite{bad} for some examples where bad currents
acquire additional renormalizations.}

After these sobering insights about formal limitations in
studying parity (and also rotational invariance) violations
within the LF framework, let us now turn to practical
limitations: the most powerful numerical techniques
for non-perturbative LF-calculations, the Lanzcos algorithm
\cite{hiller} and Monte Carlo techniques \cite{lfepmc}
are only good for
ground and low lying states (for given good quantum numbers).
Excited states and scattering states are very difficult
to handle and analyze.
This implies that we {\it cannot} (for practical reasons)
use selection rules in the decay of resonances to study
parity violations either!

\section{Parity Sensitive Observables that are Sensible}
\label{sec:sense}
The combination of these two restrictions,
exclusion of bad currents and ground state 
(for given good quantum numbers) properties only,
severely constrains the possibilities for studying
parity violations on the LF in a non-perturbative
framework, which is probably why this subject has so far not
been studied in more detail.

Fortunately, despite these limitations, there are a few observables
left which are are not excluded from the start:
Our first example is the inelastic electro-magnetic
form factor. Current conservation, i.e.
$q_\mu j^\mu=0$ and parity invariance imply that it
should be possible to write the matrix elements
of the current operator in the form
\begin{eqnarray}
\langle m,p'|j^\mu(0)| n, p \rangle
=\left\{ \begin{array}{cl}
\left( p^\mu +p'^\mu \right) F_{mn}(q^2)&
\mbox{same parity}
\\[2.ex]
q^\mu F_{mn}(q^2) & \!\!\!\!\!\!\!\!\!\mbox{opposite parity},
\end{array}\!\!\!\!\!\!\!\!\!\!\!\!\!\!\!\!\!\!\!\!
\right.
\nonumber\\[2.ex]
\label{eq:fmn0}
\end{eqnarray}
where $q=p-p'$. For the ``good'' component, this implies
\begin{eqnarray}
\langle m,p'|j^+(0)| n, p \rangle
=\left\{  \begin{array}{cl}
\left( p^+ +p'^+\right) F_{mn}(q^2)
& \mbox{same parity}
\\[2.ex]
q^+ F_{mn}(q^2) & \!\!\!\!\!\!\!\!\!\mbox{opposite parity}.
\end{array}\!\!\!\!\!\!\!\!\!\!\!\!\!\!\!\!\!\!\!\!
\right.
\nonumber\\[2.ex]
\label{eq:fmn}
\end{eqnarray}
It is implied that the states have been normalized covariantly.
Otherwise one has to multiply by appropriate normalization factors.
After factoring out the kinematic
coefficient
[$\left( p^+ +p'^+\right)$ for transitions between states of
equal parity and $q^+$ for parity changing 
transitions] the form factor should depend on
$q^2$ only, but no longer on $q^+=p^+-p'^+$ and
$q^-=M_n^2/2p^+ - M_m^2/2p'^+$ separately.
Of course if parity is violated then any
linear combination of $\left( p^+ + p'^+\right)$
and $q^+$ (with two independent form factors)
may occur on the r.h.s. and it is no longer possible 
to obtain a result that depends on $q^2$ only by
factoring out a kinematic coefficient.
Denoting $x=q^+/p^+$, energy conservation
implies
\begin{equation}
M_n^2 = \frac{q^2}{x} + \frac{M_m^2}{1-x},
\label{eq:econs}
\end{equation}
i.e. for fixed $q^2$ one obtains a quadratic
equation in $x$
\begin{equation}
x^2 M_n^2 + x\left(M_m^2-M_n^2-q^2\right) +q^2 =0,
\end{equation}
whose solutions typically come in pairs $x_{1/2}$
(except for $M_n = M_m \pm \sqrt{q^2}$).
Physically these two solutions correspond
to the two cases where the momentum is
transferred to the initial state 
by ``hitting it'' from the left or from the
right. Of course, in a parity invariant
theory one should (up to a kinematic factor)
obtain the same form factor from these two
values of $x$. However, in a LF calculation this is
in general not manifestly true and one can use
the equality of the form factor $F_{mn}$
as extracted from $x_1$ and $x_2$ as a condition
to test parity invariance. Note that this test
only works for $m\neq n$ because for $m =n$
invariance under PT (which is usually manifest
in the LF formalism), combined with longitudinal
boost invariance, guarantees equality of the
two form factors extracted from $x_1$ and $x_2$.

When one wants to test whether some LF Hamiltonian
gives rise to a parity invariant theory, one can
use the following procedure:
\begin{itemize}
\item One diagonalizes the Hamiltonian and determines
the meson wave functions
\item Then
one calculates the inelastic transition form factor
$F_{mn}$ using Eq.(\ref{eq:fmn}) as a function of
the longitudinal momentum transfer $x$ for two
arbitrary meson states $m$ and $n$\footnote{In practice one
preferably calculates the form factor between the
two lightest mesons since for those the numerical
convergence is fastest).}.
\item Then one also calculates the invariant
momentum transfer $q^2$ also as a function of $x$ using
Eq.(\ref{eq:econs})
\item Finally, one parametrically (parameter $x$) plots
$F_{mn}$ versus $q^2$. If $F_{mn}$ does not turn out to be
a unique function of $q^2$ then parity is violated
\end{itemize}
Below the practicality of this procedure will be demonstrated
in a concrete example. However, before doing this, we should
mention a possible caveat:
In order to be able to evaluate the matrix element appearing
in Eq.(\ref{eq:fmn}) one needs two ingredients:
the states in some basis and the current operator in the
same basis. For the ``good'' current appearing in Eq.(\ref{eq:fmn})
we will always assume the canonical form in this paper.
There are several reasons for doing this. First, it is the most
simple form. Second, since no Tamm-Dancoff approximation will
be employed in this paper and since we extrapolate to
the limit where there is no cutoff in this paper, there is
no reason to believe that the good current should be modified
from its canonical form.\footnote{In this limit, zero-mode
corrections are expected only for bad currents!} 
Third, our numerical results explicitly
demonstrate that this leads to a self-consistent procedure.
Nevertheless, especially when employing Tamm-Dancoff truncations,
or other parity violating (un-extrapolated) cutoffs, one must
consider modifying even the good current operator from its
canonical form. A detailed discussion of this would go beyond
the intended scope of this paper, but it should be
emphasized that parity conditions might also be helpful
in a self-consistent determination of the current operator in those
cases.

\section{A non-perturbative example}
\label{sec:nonex}
The most simple example, where the issue of parity invariance
and LF quantization can be studied, is the Yukawa model in
1+1 dimensions
\begin{equation}
{\cal L} = \bar{\psi} \left( i \not \! \partial
-M -g\phi \right) \psi + \frac{1}{2}\partial_\mu \phi
\partial^\mu \phi - \frac{m^2}{2} \phi^2,
\label{eq:yuk}
\end{equation}
where $\phi$ is some scalar field.
The LF quantization of this model has been studied in Refs.
\cite{pa:dlcq,yan,hari:yuk,dallasmafia,osu:nonloc}.
With $\psi_{(\pm )}$ as defined above, the
spinor part of the above Lagrangian (\ref{eq:yuk}) reads\cite{yan}
\begin{eqnarray}
{\cal L}_\psi &\equiv& \bar{\psi} \left( i \not \! \partial -M-g\phi
\right) \nonumber\\
&=& \sqrt{2} \left[ \psi_{(+)}^\dagger i \partial_+ \psi_{(+)}
+ \psi_{(-)}^\dagger i\partial_- \psi_{(-)}\right]\nonumber\\
& &- \left(M+g\phi\right)\left[\psi_{(+)}^\dagger \gamma^0 \psi_{(-)}
+\psi_{(-)}^\dagger \gamma^0\psi_{(+)}\right]
\label{eq:lpsi}
\end{eqnarray}
and $\psi_{(-)}$ satisfies the constraint equation
\begin{equation}
\psi_{(-)} = -\frac{i}{\sqrt{2} \partial_-} \left(M+g\phi\right)
\gamma^0 \psi_{(+)}. 
\label{eq:constr}
\end{equation}
Upon inserting the solution to this constraint equation (\ref{eq:constr})
into ${\cal L}_\psi$ (\ref{eq:lpsi}) one finds
\begin{eqnarray}
{\cal L}_\psi &=& \sqrt{2}\psi_{(+)}^\dagger i\partial_+\psi_{(+)}
+ \frac{M^2}{\sqrt{2}} \psi_{(+)}^\dagger \frac{i}{\partial_-}\psi_{(+)}
\label{eq:lpsi2}\\
&+& \frac{Mg}{\sqrt{2}} \psi_{(+)}^\dagger \!\!\!\left[
\frac{i}{\partial_-} \phi + \phi\frac{i}{\partial_-}\right] \!\!\psi_{(+)}
+ \frac{g^2}{\sqrt{2}} \psi_{(+)}^\dagger \phi \frac{i}{\partial_-} 
\phi \psi_{(+)},
\nonumber
\end{eqnarray}
which contains only dynamical degrees of freedom and can
be quantized straightforwardly. One thus obtains the {\it canonical}
LF Hamiltonian 
\begin{equation}
P^-_{can}=\int dx^- {\cal H}_{can},
\label{eq:hyuk1}
\end{equation}
where
\begin{eqnarray}
{\cal H}_{can} \!&=& \frac{m^2}{2}\phi^2
- \frac{M^2}{\sqrt{2}} \psi_{(+)}^\dagger \frac{i}{\partial_-}\psi_{(+)}
\label{eq:hyuk2}\\
&-& \frac{Mg}{\sqrt{2}} \psi_{(+)}^\dagger \!\!\!\left[
\frac{i}{\partial_-} \phi + \phi\frac{i}{\partial_-}\right] \!\!\psi_{(+)}
- \frac{g^2}{\sqrt{2}} \psi_{(+)}^\dagger \!\phi \frac{i}{\partial_-}
\phi \psi_{(+)}.
\nonumber
\end{eqnarray}
The canonical Hamiltonian (\ref{eq:hyuk1}) contains 4 terms:
a two point function for both fermions and bosons, a three point
function and a four point function. From the point of view of
renormalization it is thus natural to make the following ansatz
for the renormalized Hamiltonian density\footnote{In fact, 
in perturbation theory, it is both necessary
but also sufficient to generalize the Hamiltonian 
as in Eq.(\ref{eq:hyuk3}) \cite{mbrot}.}
\begin{eqnarray}
{\cal H}_{ren} &=& \frac{m^2}{2}\phi^2
- \frac{M^2}{\sqrt{2}} \psi_{(+)}^\dagger \frac{i}{\partial_-}\psi_{(+)}
+{\cal H}_{n.o.}
\label{eq:hyuk3}\\
&-& \frac{c_3}{\sqrt{2}} \psi_{(+)}^\dagger \!\!\!\left[
\frac{i}{\partial_-} \phi + \phi\frac{i}{\partial_-}\right] \!\!\psi_{(+)}
- \frac{c_4}{\sqrt{2}} \psi_{(+)}^\dagger \phi \frac{i}{\partial_-} 
\phi \psi_{(+)}.
\nonumber
\end{eqnarray}
The canonical Hamiltonian is obtained by taking \cite{pa:dlcq}
\begin{equation}
c_3 = \sqrt{M^2c_4}\quad \quad \mbox{(canonical Hamiltonian)}.
\label{eq:canrel}
\end{equation}
In perturbation theory with ${\cal H}_{can}$, infinities in the
longitudinal momentum integral occur only at the one-loop level
(for both fermion and boson self-energies)
and are calculable\cite{pa:dlcq}. The corresponding counter term (whose
infinite part is unique) is denoted by the ``normal ordering
term'' ${\cal H}_{n.o.}$. The finite part of ${\cal H}_{n.o.}$
has the operator structure of kinetic terms.
Since such operators are already explicitly included in the
above ansatz [Eq.(\ref{eq:hyuk3})], it is not necessary to
discuss the finite part of ${\cal H}_{n.o.}$ here any further.

The Lagrangian as well as the canonical Hamiltonian
(\ref{eq:hyuk1}) contain only
3 free parameters:$ M,m,g$. Therefore the most general situation for
the Yukawa model should thus be described by fixing only
3 parameters as well. On the other hand, it is known
\cite{mbrot,mb:adv} that the above relation (\ref{eq:canrel})
is not valid after renormalization, i.e. it seems that
all 4 parameters in Eq.(\ref{eq:hyuk3}) get renormalized
independently. However, the 4 parameters in Eq.(\ref{eq:hyuk3})
are not really independent! The point is that arbitrary values
for the parameters $m^2,\, M^2,\, c_2,\, c_4$ do {\it not}
correspond to the Yukawa model but rather something else.
Only for a 3-dimensional subspace of the 4-dimensional
parameter space spanned by $m^2,\, M^2,\, c_2,\, c_4$ does
Eq.(\ref{eq:hyuk3}) actually describe a version of the
Yukawa model. At the tree level, this 3-dimensional
subspace is characterized by the canonical relation 
(\ref{eq:canrel}). Beyond the tree level, the relation
between the 4 parameters, for which Eq.(\ref{eq:hyuk3})
describes a Yukawa model, is in general more complex.
Hence the crucial question is: how can one find that
relation? One possibility (at least in principle)
is that one makes calculations
both using equal time quantization as well as LF quantization.
One then calculates 4 physical quantities in both schemes
and fine-tunes the 4 parameters in the LF calculation
until the 4 physical quantities have been reproduced.
Even though there is nothing fundamentally wrong with this
procedure, it is very unattractive since it requires one
to go back to an equal time quantized theory in order to
define the LF theory.

A much more satisfying approach is based on the observation
that, in general, a ``wrong'' combination of the 4 parameters
$m^2,\, M^2,\, c_2,\, c_4$ leads to a parity violating theory!
\footnote{One can easily convince oneself about this fact 
for example by calculating some physical observables at tree level,
but with a combination of parameters that does not satisfy
the canonical relation [Eq.(\ref{eq:canrel})].}
One can exploit this fact by means of the following renormalization
procedure: first one picks (or fixes, using some at this point
unspecified renormalization condition)
three of the above four parameters. Then one fine tunes the
fourth parameter until some parity sensitive observable
indicates no violation. As a consistency condition one can
check more than one parity sensitive observable.

In the application to the Yukawa model we used a variation of this
generic procedure. First, the four point coupling $c_4$
is assigned the value $2\pi$.
\footnote{Note, Yukawa$_{1+1}$ is superrenormalizable and there
is only a finite renormalization of the four point coupling
which we ignore here for simplicity. Alternatively we could have
imposed a condition on a physical observable here.}
This only determines the
overall mass scale. A completely equivalent approach would have
been not to fix $c_4$ at all but to measure all dimensionful
parameters and physical quantities in units of $\lambda\equiv c_4/2\pi$ 
(which carries the dimension mass squared).

Then arbitrary values for the physical masses of the lightest
fermion as well as the lightest scalar meson (strictly speaking,
the $C=1$ meson, which is supposed to be scalar!) were selected.
Including the condition $c_4=2\pi$, this implies 3 conditions
and we can now start the actual fine tuning procedure which
allows to fix all four bare parameters:
For given values of $M^{phys}_F$ and $M^{phys}_S$
an arbitrary value for the vertex mass (i.e. for $c_3$)
was selected.
Then the bare masses of both fermion and boson were
fine tuned so that the physical masses of both fermion
and boson equal $M^{phys}_F$ and $M^{phys}_S$.
After this had been achieved, the inelastic
transition form factor between the two lightest mesons
was calculated as described above. This procedure was repeated for
different values of $c_3$ until the inelastic
form factor satisfied the parity condition discussed in section
\ref{sec:sense}.

\begin{figure}
\unitlength1.cm
\begin{picture}(10,15.4)(2.6,1.2)
\includegraphics{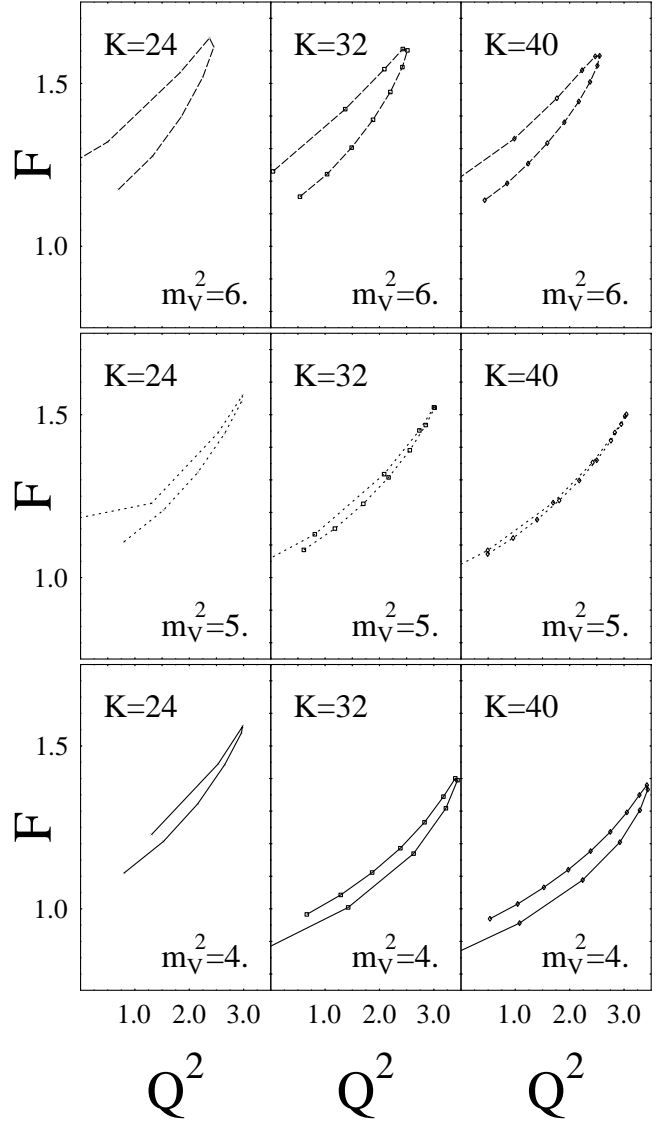}
\end{picture}
\caption{
Inelastic transition form factor (\protect\ref{eq:fmn}) between the
two lightest meson states of the Yukawa model, calculated
for various vertex masses $m_v$ and for various DLCQ parameters
$K$. The physical masses for the fermion and the scalar meson
have been renormalized to the values 
$\left(m_F^{phys}\right)^2=\left(m_F^{phys}\right)^2=4$.
All masses and momenta are in units of
$\protect\sqrt{\lambda} = \protect\sqrt{c_4/2\pi}$.
}
\label{fig1}
\end{figure}

\begin{figure}
\unitlength1.cm
\begin{picture}(10,15.4)(2.6,1.2)
\includegraphics{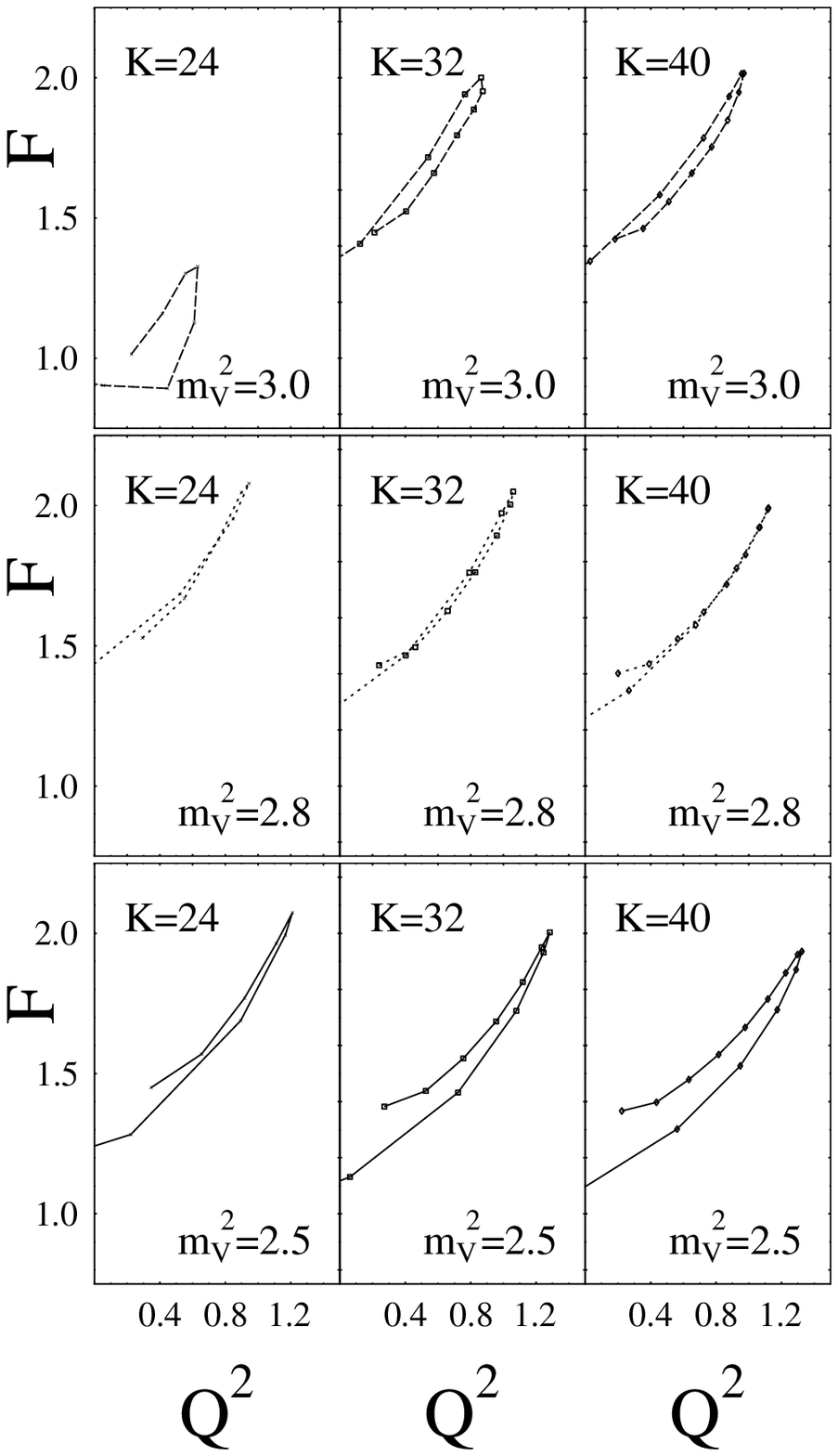}
\end{picture}
\caption{Same as Fig.\protect\ref{fig1}
but for $\left(m_F^{phys}\right)^2=\left(m_F^{phys}\right)^2=2$.
}\label{fig2}
\end{figure}

In the numerical procedure, DLCQ \cite{pa:dlcq}, 
with anti-periodic boundary conditions for the fermions
and periodic boundary conditions for the bosons, was employed.
No cutoff beyond the DLCQ cutoff was used, i.e.
particle number was allowed to reach arbitrarily
large values --- the only limit was set by the DLCQ
parameter $K$, which measures the total momentum of the initial
meson in the transition matrix element [Eq.(\ref{eq:fmn})] in 
units of $2\pi/L$, where L is the box length in DLCQ.
The DLCQ-cutoff itself violates parity.
It is therefore necessary, to verify that the results
have numerically converged in $K$. 
The matrix elements were calculated for
$K=24,\, 32,\, \mbox{and} \, 40$. Typical results
are shown in Figs. \ref{fig1} and \ref{fig2}.
Fig. \ref{fig1} corresponds to intermediate coupling
[$\left(M_F^{phys}\right)^2=\left(M_F^{phys}\right)^2= 4$
in units of the coupling constant], while Fig.\ref{fig2}
[$\left(M_F^{phys}\right)^2=\left(M_F^{phys}\right)^2= 2$]
is an example for strong coupling.
\noindent
In the calculations, equal physical masses for the fermion 
and the scalar meson were chosen because if the fermion
and meson have similar masses there is only one scale and
the numerical convergence (in $K$) is faster. In principle,
there is nothing wrong with repeating this procedure for
unequal masses.

Since momenta assume only discrete values in DLCQ, the
form factor could only be evaluated at a discrete set of points.
For example, with an initial momentum of $K=40$ for the
``pseudoscalar'' meson, the final momentum of the ``scalar'' 
meson was taken to be $K=38,36,34,..$ .
In the plots, the form factors, evaluated between states with these
momenta, were connected by a smooth curve to guide the eye.

From the fact that
the form factors for $K=32$ and $K=40$ hardly differ, one can
conclude that the results have converged numerically.
The examples in Figs.\ref{fig1} and \ref{fig2} show
several facts:
\begin{itemize}
\item For arbitrary combinations of physical masses (or bare
kinetic masses) and vertex masses, the transition form factor
$F_{mn}$ (\ref{eq:fmn}) is {\it not} a unique function of
$q^2$ (even for $K$ large), thus clearly demonstrating a violation
of parity in the general case.
\item For specific combinations of physical masses (or bare
kinetic masses) and vertex masses, the transition form factor
$F_{mn}$ (\ref{eq:fmn}) is a unique function of
$q^2$. For given values for the physical masses,
a unique vertex mass, which renders the form factor
parity invariant, was found: In the case of
$\left(m_F^{phys}\right)^2=\left(m_F^{phys}\right)^2=4$,
the correct value for the vertex mass is about $m_V^2 \approx 5$,
while for $\left(m_F^{phys}\right)^2=\left(m_F^{phys}\right)^2=2$
the correct value is near $m_V^2 \approx 2.8$.
\item It should be emphasized that the transition
form factor is a function and not just one number, i.e.
the mere fact that $F_{mn}$ is parity invariant over the
whole range of $q^2$ considered provides a consistency
check of the procedure described in this work.
This fact, plus the uniqueness mentioned above, give a strong
indication that we have really found the correct renormalized
Hamiltonian and that the procedure outlined in this paper is
practical.
\end{itemize}

As a side remark, it should be explained here how the bare
masses were fixed in practice. There are two slightly different
procedures one can imagine. 
In the first procedure one first adds a momentum
dependent kinetic term that takes care of the one loop
divergences in the self-energies. Then one adds
a finite (momentum independent) bare masses for boson
and fermion which are fine tuned until the physical masses
take the values desired in the large $K$ limit. 
In this work, a slightly different procedure was used:
momentum dependent bare masses for both boson and fermion were 
introduced in such a way
(by fine-tuning) that the physical masses are $K$ independent.
This can be easily done in a successive procedure.
In the large $K$ limit, the momentum dependence of the
kinetic term thus obtained
reproduces the momentum dependence as derived from the
one loop counter term and therefore both procedures agree
with each other in the large $K$ limit. However, it was found
that, typically, when the physical masses of the lightest
particles are exactly $K$
independent (and not only in the limit $K\rightarrow \infty$),
other physical observables converge faster to their
$K\rightarrow \infty$ values.

\section{Summary}
\label{sec:summary}
We have investigated several classes of observables, which
are potentially sensitive to parity violations, and found
that most of them are {\it not} suitable for ``typical''
LF calculations. What makes them ``unsuitable'' is
that they involve scattering states (which are notoriously
difficult for most non-perturbative numerical algorithms) or they involve
matrix elements of
``bad'' currents (which are often ill defined in the LF
formalism). Other observables, which seem {\it a priory}
sensitive to parity violations of the formalism are often
``protected'' by manifest LF symmetries, such as C-parity.

We found one observable which is both sensitive to parity
violations and easily accessible in a standard LF calculation:
inelastic matrix elements of the good component of the
current operator.
Vector current conservation demands
\begin{equation}
q^-\langle m,p'|j^+(0)| n, p \rangle
+q^+
\langle m,p'|j^-(0)| n, p \rangle
=0,
\label{eq:par1}
\end{equation}
while parity invariance implies
\begin{equation}
\langle m,p'|j^-(0)| n, p \rangle = \mp
\langle m,\bar{p'}|j^+(0)| n, \bar{p} \rangle,
\label{eq:par2}
\end{equation}
where $\bar{p}^\pm \equiv p^\mp =M_n^2/2p^\pm$, 
$\bar{p'}^\pm \equiv p'^\mp=M_m^2/2p'^\pm$ are
the parity transformed momenta of the initial and
final state
and the sign in Eq.(\ref{eq:par2}) depends on
whether the two states $m$ and $n$ have the same
or opposite intrinsic parity.
Eqs.(\ref{eq:par1}) and (\ref{eq:par2})
individually are useless, since they involve a
bad current matrix element.
However, notice that the bad current matrix element
in Eq. (\ref{eq:par1}) and in Eq. (\ref{eq:par2})
is the same. The trick is to use Eq. (\ref{eq:par2}) to eliminate
the matrix element of $j^-$ from
Eq.(\ref{eq:par1}) and one obtains a relation between
two matrix elements of the good current at different
values of $q^+$ but at the same value of $q^2$, yielding
\begin{equation}
q^-\langle m,p'|j^+(0)| n, p \rangle
=\pm q^+
\langle m,\bar{p'}|j^+(0)| n, \bar{p} \rangle
.
\label{eq:par3}
\end{equation}
We demonstrated explicitly that the parity relation thus obtained
(\ref{eq:par3}) is useful for practical calculations by applying it 
to a concrete non-perturbative example: Yukawa$_{1+1}$. In the LF 
formulation, the renormalized Hamiltonian contains one more ``free''
parameter than the Lagrangian. Since the additional counter term
involved is not parity invariant one can use the parity
relation derived in this paper as an additional renormalization
condition. We studied a few cases numerically and showed that
the fine tuning procedure can be done also in practice.
Since the parity relation for the inelastic transition
form factor is in fact not just one relation, but an infinite
number (it involves functions!), we obtained at the same time
a strong self consistency check for the whole procedure.

Can a similar procedure be applied to help in constructing the 
LF-Hamiltonian for QCD? First of all, since the parity relation
derived in this paper involves only meson states, color singletts
and physical (gauge invariant) operators, there is nothing special
about QCD. Even though QCD is a gauge theory and even though one
usually picks the $A^+=0$ (or similar) condition on the gauge field,
the parity relation for the good current should still hold
if the LF formulation is to be parity invariant at the level of
physical observables. However, some conditions must be satisfied before 
one can do this in practice: Either one must be sure that all cutoffs that
violate parity, such as Tamm-Dancoff truncations of longitudinal
momentum cutoffs or energy cutoffs must be taken large 
(or small) enough so that they no longer affect the states under
consideration. This is hard, but not impossible! For example,
on a small transverse lattice, one can get rather close to the 
longitudinal continuum limit in practical calculations.
However, in examples were one cannot remove those cutoffs, one must carefully
renormalize the currents before calculating the necessary matrix 
elements. 

Even though we focused on inelastic transition matrix elements of
the good component of the vector current,
there are more ``useful'' parity relations beyond the inelastic
vector transition matrix element. For example, there exist 
also some useful ``parity-relations'' among virtual Compton 
amplitudes which may also be useful in the context of QCD. 
However, a detailed investigation of such observables 
will be postponed to some forthcoming paper.

In any case, parity relations such as the ones derived in this
paper should be useful in the search for $P^-_{QCD}$, since they
imply a strong consistency check for the states that arise
as solutions from diagonalizing $P^-_{QCD}$.

%\begin{figure}
%\caption{
%Inelastic transition form factor (3.2)
%between the
%two lightest meson states of the Yukawa model, calculated
%for various vertex masses $m_v$ and for various DLCQ parameters
%$K$. The physical masses for the fermion and the scalar meson
%were renormalized to the values 
%$\left(m_F^{phys}\right)^2=\left(m_F^{phys}\right)^2=4$.
%}
%\label{fig1}
%\end{figure}

\end{document}